\documentclass[3p,times]{elsarticle}
\usepackage{amssymb}
\usepackage{amsthm}
\usepackage{amsmath}

\usepackage{epstopdf}

\usepackage{graphicx}              
\usepackage[figuresright]{rotating}
\usepackage{subcaption}

\begin{document}

\begin{frontmatter}




\title{Afterpulsing in Silicon Photomultipliers: 	Impact on the Photodetectors Characterization}


\author{Adam Para}

\address{Fermi National Accelerator Laboratory, Batavia, IL60510,USA}

\begin{abstract}
Novel generation of silicon-based photodetectors are attractive alternatives to the traditional phototubes. They offer significant
advantages but they present new challenges too.   Presence of afterpulses may affect many characteristics of the photodetectors. Simple statistical model of afterpulsing is used to evaluate the contribution to the observed dark count rates, to examine the contribution to the pulse height resolution and to demonstrate the modification of the observed timing properties of the SiPMs.

\end{abstract}



\end{frontmatter}


\section{Introduction} \label{intro}
Silicon photomultipliers (SiPM's, MPPC's) are the arrays of avalanche photodiodes operating in a Geiger mode. Their high gain,
 high detection efficiency and compact sizes in conjunction with very high pulse height resolution enabling 'photon counting' make them very attractive
candidates for replacement of the traditional photomultipliers, PMTs, especially in applications where the total area of
photodetectors required is small.

SiPM's offer several advantages over the PMT tubes but they also present new challenges related to  solid state
nature of these devices. Thermally generated free charge carriers create signals identical to those caused by photoelectrons,
hence the dark count rates in typical devices are relatively large, in the range $10^4cps/mm^2-10^6cps/mm^2$\cite{Hamamatsu}.

 Large number of carriers are present in the Geiger discharge; some of them may be trapped in metastable traps. Their subsequent
release may produce additional pulses, afterpulses, which contribute to the observed signal. The stochastic nature of this
process affects the performance of photodetectors.

 Afterpulsing is a result of a large number of independent 'events' (electron traversing the junction) hence it can be well modelled using standard statistical 
methods. In principle one can derive any observable quantity or distribution by a suitable convolution of Poisson and exponential 
distributions with the mean values and lifetimes appropriate for a detector in question. We have calculated these convolutions using Monte Carlo integration and generating 50 million initial afterpulses.

\section{Modeling of  Afterpulses; Single Pulse} \label{mode_singlel}

Afterpulses can be characterized in terms of the probability that a given pulse will produce a subsequent  afterpulse and their time constant $\tau_{aft}$. In this note  the  $P_{aft}$ will be used to denote the average number of  trapped electrons (hence the number of afterpulses produced by a single pulse, not counting the next generation of afterpulses).
 For  $P_{aft} \ll 1$ it has the meaning of the actual probability of producing an afterpulse.  These
parameters depend on the operating point of the SiPM. Increase of the bias voltage $V_{bias}$ leads to corresponding increase
of the  $P_{aft}$ because of the increase of the number of charge carriers in the avalanche
$Q=C_{cell}(V_{bias}-V_{breakdown})=C_{cell}V_{ov}$, where $C_{cell}$ is the cell capacitance and the $V_{breakdown}$ is the breakdown voltage,
and the simultaneous increase of the probability that a free charge carrier will initiate an avalanche, $P_{Geiger}$.
The afterpulsing time constant  $\tau_{aft}$ may depend on the temperature of the photodetector.

In general there might be several trapping levels present with different trapping probabilities and corresponding time constants.
In the following we explore a simple model of one trapping level with a single afterpulsing probability and one time constant.

Afterpulsing is a stochastic process, the actual number of additional pulses undergoes fluctuations according to the Poisson
distribution with  $P_{aft}$ being the expectation value. In the note the quantity $P_{aft}$ will be referred to as 'afterpulsing
probability'.   Every afterpulse produces a standard avalanche, hence it may lead to
a production of additional afterpulses.

In a realistic detector the situation is somewhat more complicated as the bias voltage after an avalanche drops to $V_{breakdown}$
and  is being restored with
the time constant $\tau_{RC}$ characteristic for a given device. Electrons released from the traps at timescales shorter
than $\tau_{RC}$ have lower
probability of producing an avalanche because of reduced $P_{Geiger}$ and the charge of the avalanche will be reduced proportionally to the actual value of overvoltage,
 $V_{ov}$. The latter effect will lead to a reduced afterpulsing probability  $P_{aft}$ of the subsequent afterpulses regardless
of their timing. These effects depend on the specific devices and they are ignored in the present study.

Afterpulsing is simulated assuming a single avalanche creating a Poisson-distributed number of afterpulses distributed
with the exponential decay time. The procedure is iteratively applied to each of the generated afterpulses, thus leading to
trains of pulses with varying numbers of pulses and their time distributions.

\section{Afterpulsing Pulse Rates; Effective Gain} \label{rates}
\begin{figure}[ht]
   \begin{minipage}[b]{0.45\linewidth}
      \includegraphics[clip, width=\linewidth]{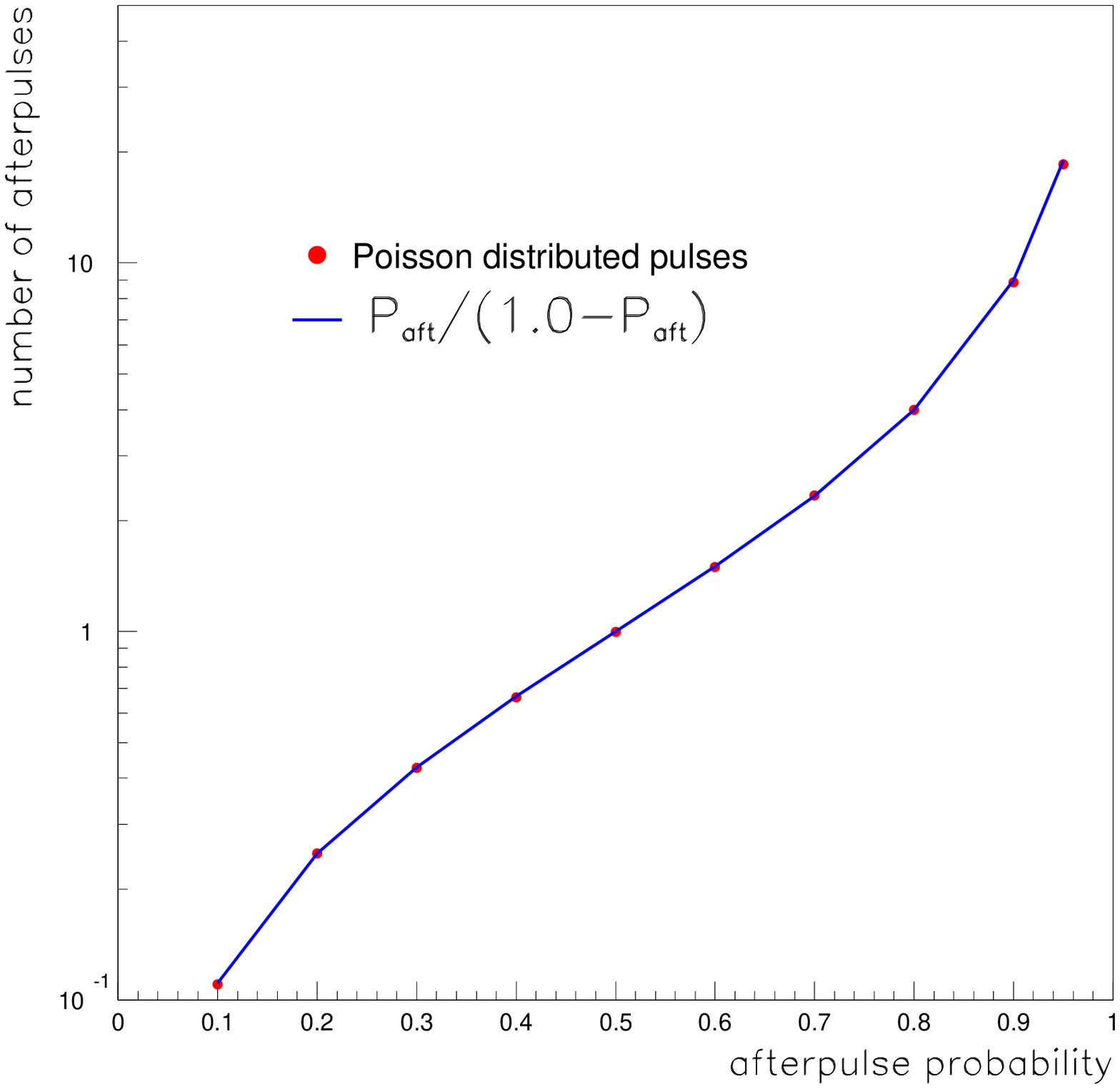}
      \caption{Average number of afterpulses as a function of the afterpulsing probability with the actual number of afterpulses undergoing statistical fluctuations.}
      \label{fig:no_of_afterpulses}
   \end{minipage}
   \quad
   \begin{minipage}[b]{0.45\linewidth}

      \includegraphics[clip, width=\linewidth]{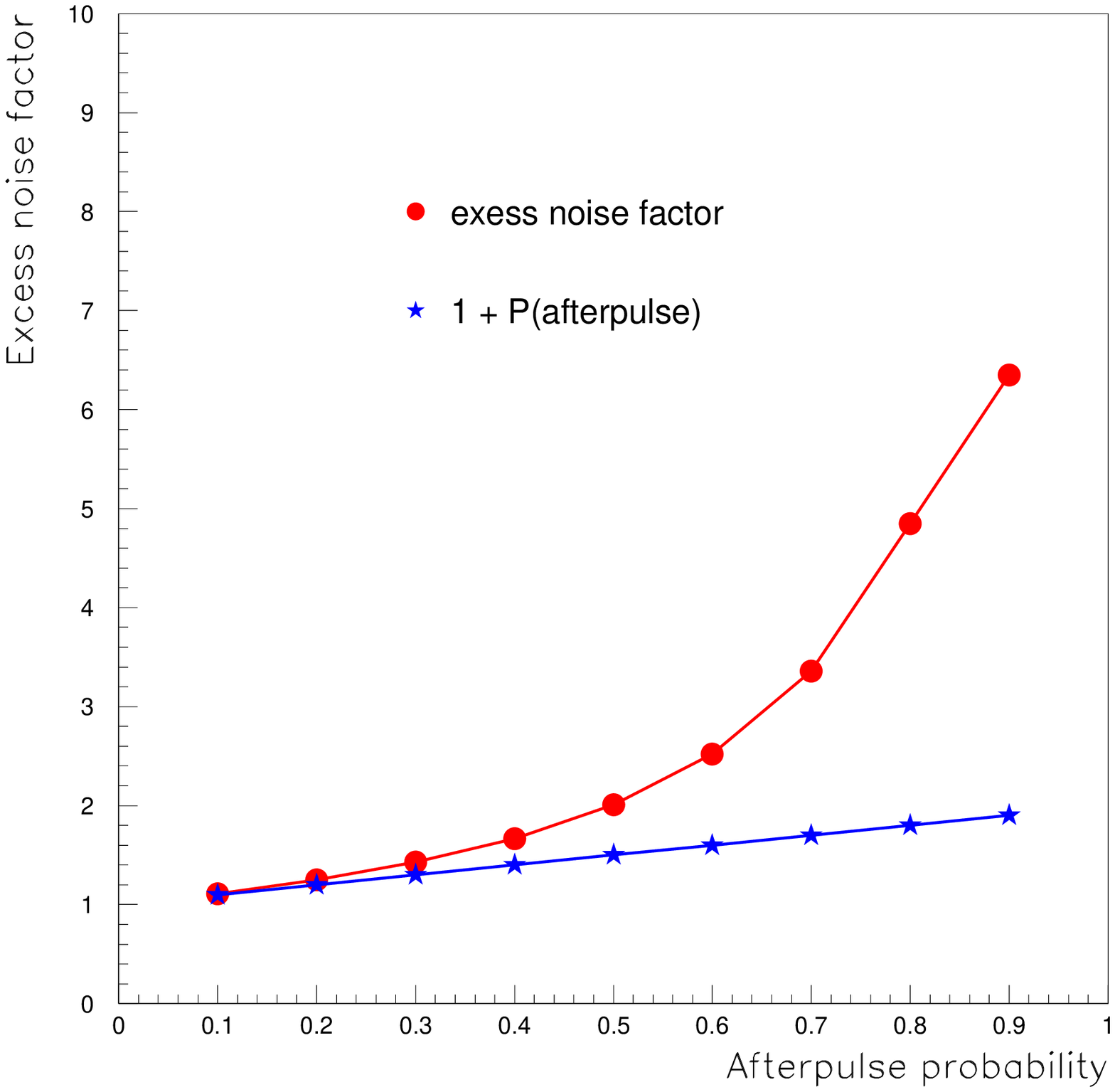}
      \caption{Excess noise factor as a function of the afterpulsing probability. For comparison a linear function $1+P_{aft}$
               is shown as a blue line.}
      \label{fig:ENF}
   \end{minipage}

\end{figure}

Afterpulsing with fixed probability would lead to an increase of the number of pulses by an additional contribution of
\begin{equation} \label{eq:No_of_afterpulses}
N_{aft} =  P_{aft} + P_{aft}^2 + P_{aft}^3 + ... =\frac{P_{aft}}{1-P_{aft}}
\end{equation}
and it may be treated as an additional gain factor, albeit dependent on the bias voltage. As shown in Fig.\ref{fig:no_of_afterpulses} the same is true even in the case when the afterpulses are created with fluctuations governed by the Poisson statistics.

The additional gain factor undergoes additional fluctuations thus it reduces the intrinsic resolution of the photodetector,
which can be expressed in terms of the Excess Noise Factor, $ENF$\footnotemark[1]. Fig. \ref{fig:ENF} shows that the $ENF$, 
calculated from the mean value and the RMS of the distribution of the number of afterpulses,  grows very rapidly,
much faster than linearly,  with the $P_{aft}$.

\footnotetext[1]{Excess noise factor is defined as $ENF=\frac{\sigma ^2}{mean}$. For  purely poissonian fluctuations $ENF=1.0$. Any deviation of $ENF$
from unity indicates additional fluctuations beyond purely statistical ones.}
The increase of the number of additional pulses with  $P_{aft}$ (thus with $V_{bias}$) is the primary reason for the observed
increase of the dark count rates with the bias voltage.
\section{Afterpulses Time Distribution} \label{time}
Afterpulsing time constant can be determined experimentally by studying the observed time distribution of pulses following some
'trigger' pulse. An external light pulse can be used to provide the initial trigger and the contribution of thermal pulses
can be subtracted. The shape of resulting time distribution is modified by the statistical nature of afterpulsing, however.

\begin{figure}[ht]
   \begin{center}
   \begin{minipage}{0.4\linewidth}
      \includegraphics[clip, width=\linewidth]{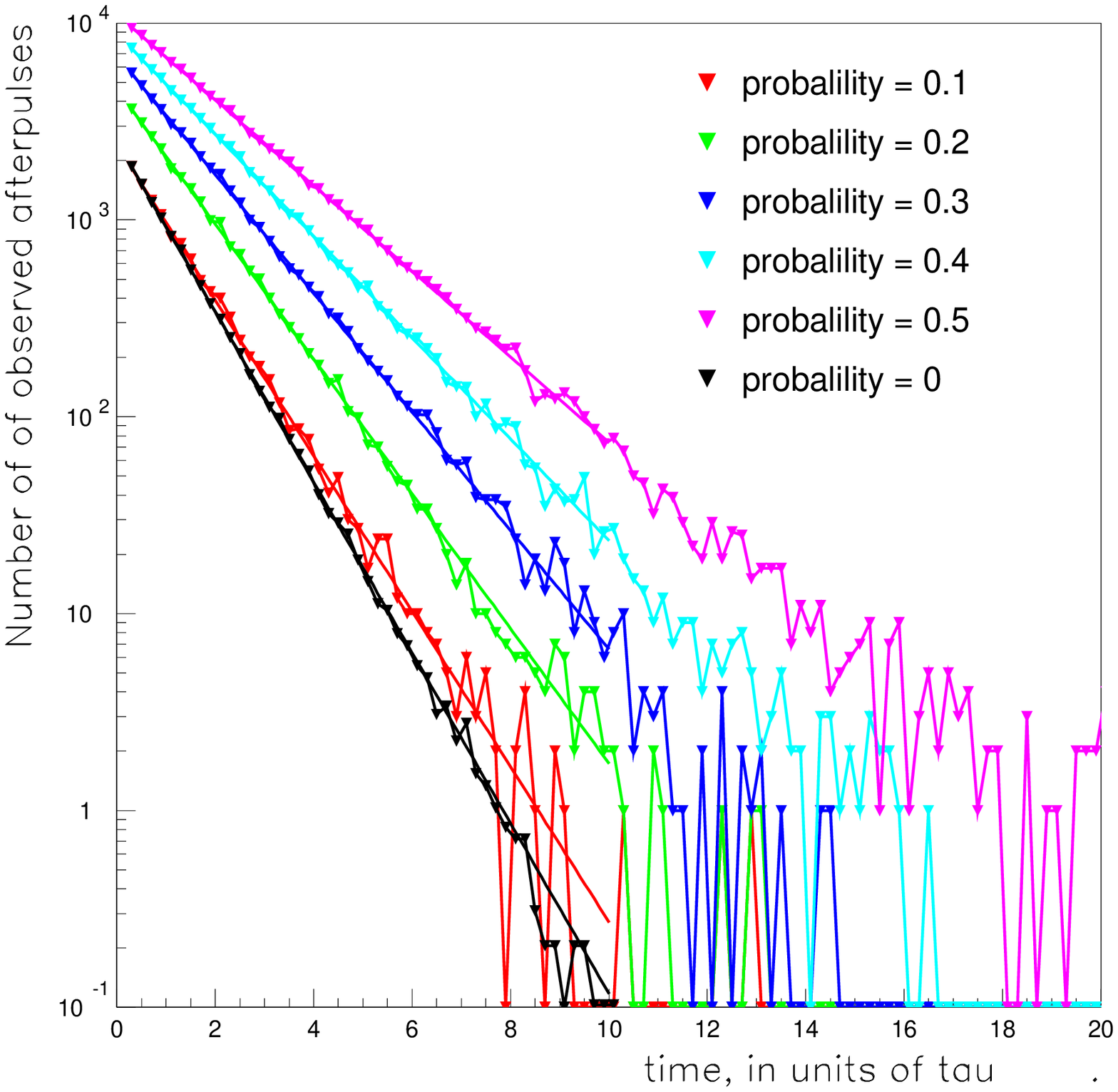}
      \caption{Time distribution, in units of $\tau_{RC}$, of the observed pulses for afterpulses probability ranging from 0.1 to 0.5. The black distribution corresponds to the underlying exponential  and no subsequent afterpulsing.}
      \label{fig:time_low_afterpulsing}
   \end{minipage}
   \quad
   \begin{minipage}{0.4\linewidth}
      \includegraphics[clip, width=\linewidth]{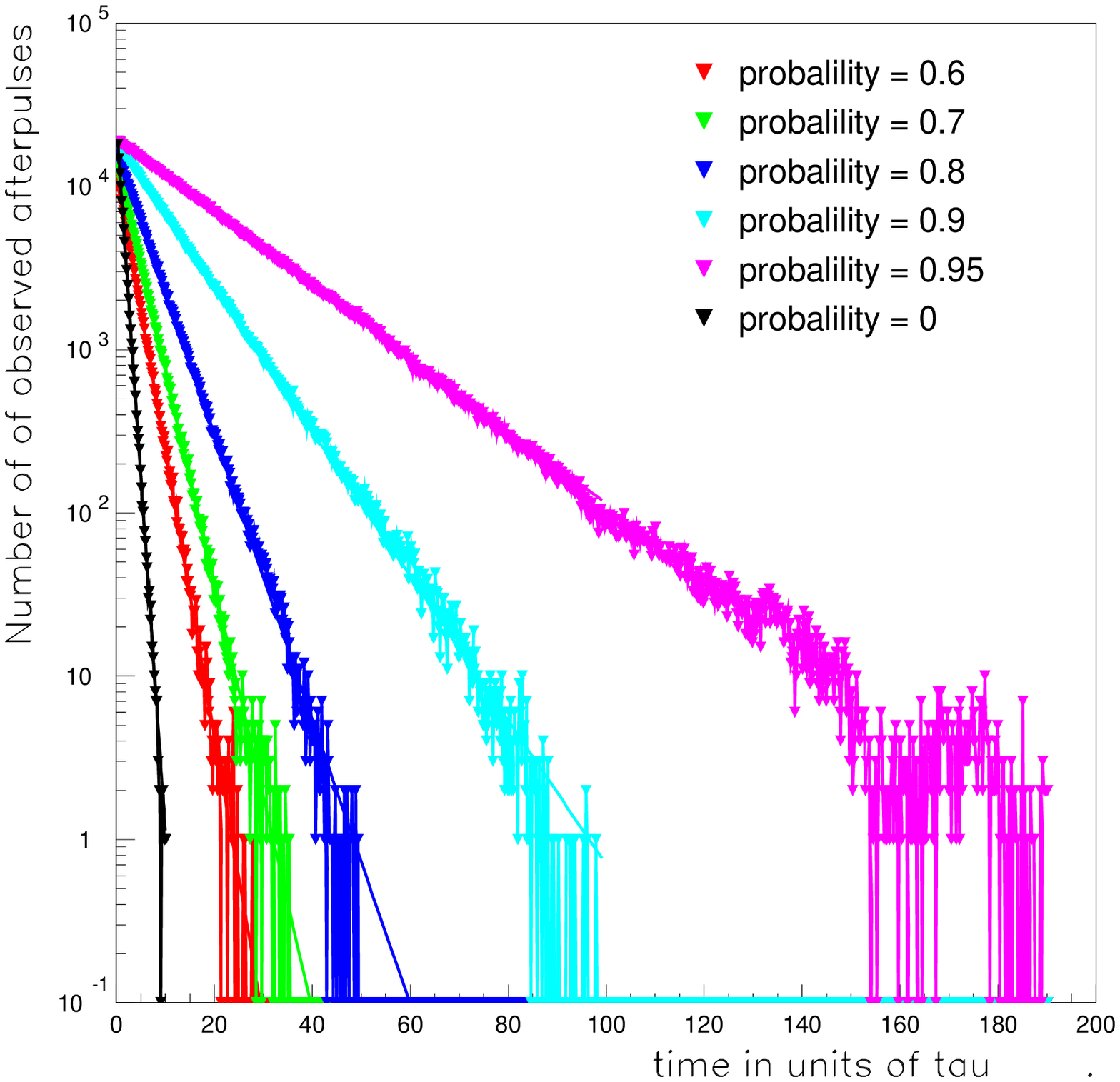}
       \caption{Time distribution, in units of $\tau_{RC}$, of the observed pulses for afterpulses probability ranging from 0.6 to 0.95. The black distribution corresponds to the underlying exponential   and no subsequent afterpulsing.}

      \label{fig:time_high_afterpulsing}
   \end{minipage}
   \end{center}
\end{figure}

\begin{figure}[hb]
\begin{center}
   \begin{minipage}[b]{0.43\linewidth}
      \includegraphics[clip, width=\linewidth]{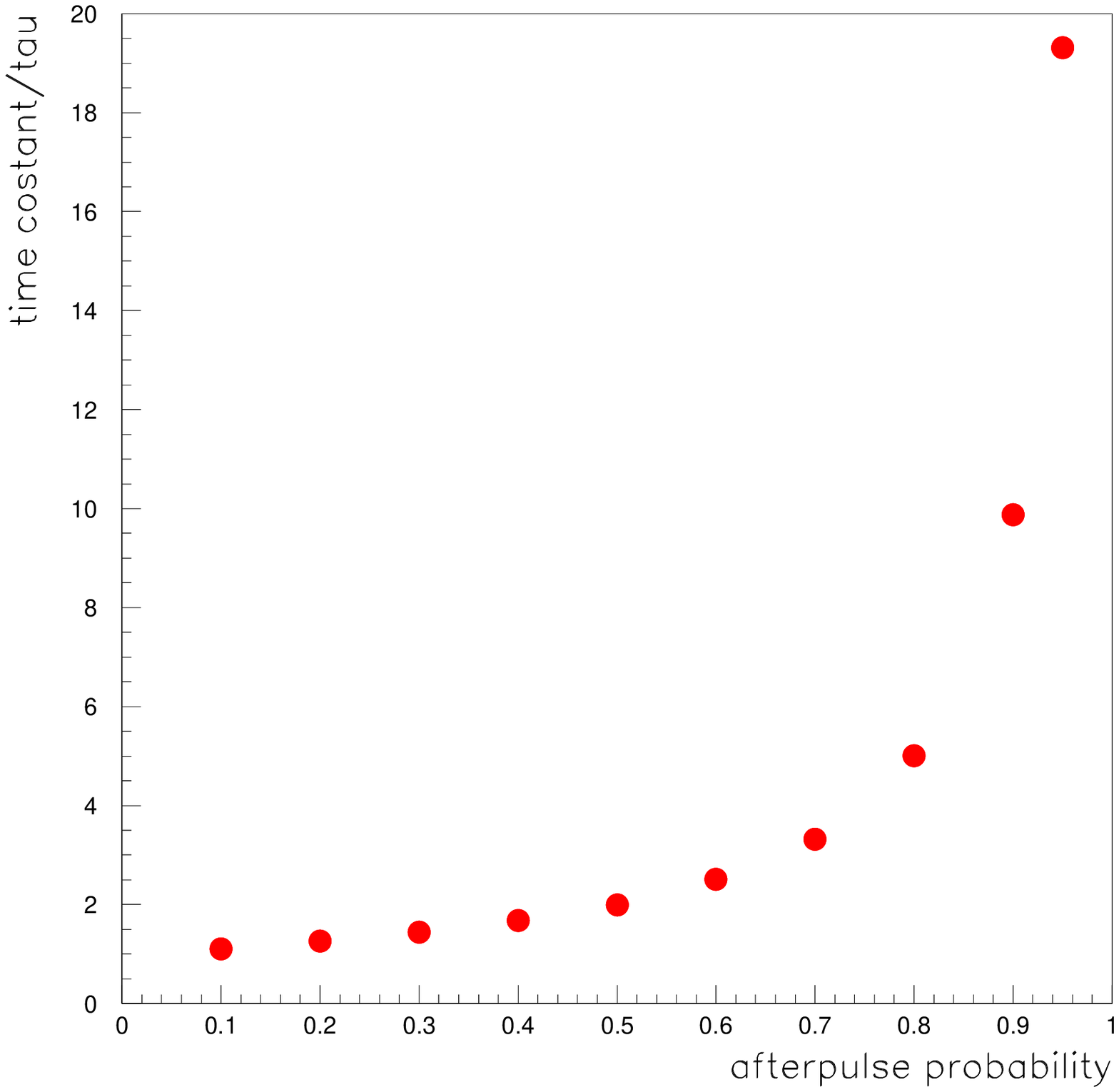}
      \caption{Effective afterpulsing time constant, in units of $\tau_{RC}$, as a function of the afterpulsing probability.}
      \label{fig:afterpulsing_time_constant}
   \end{minipage}
   \quad
   \begin{minipage}[b]{0.43\linewidth}
      \includegraphics[clip, width=\linewidth]{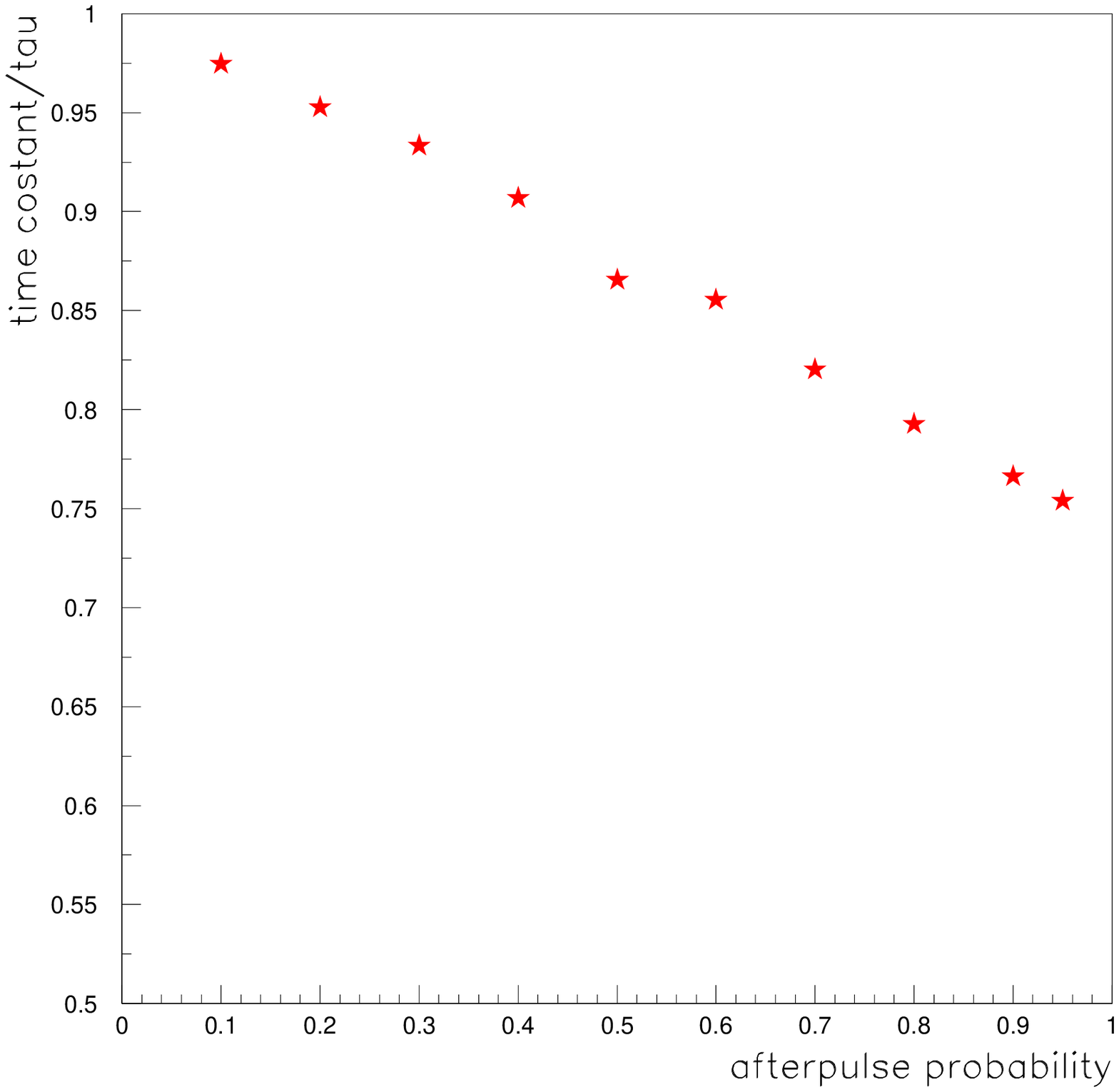}
      \caption{Time constant, in units of $\tau_{RC}$, determined from  the 'first afterpulse' as a function of the afterpulsing probability}
      \label{fig:first_afterpulse}
   \end{minipage}
\end{center}
\end{figure}

Figs. \ref{fig:time_low_afterpulsing} and  \ref{fig:time_high_afterpulsing} illustrate the impact of subsequent afterpulsing on the
 time distribution of the observed pulses for
 different values of the $P_{aft}$. The time distribution of the observed pulses retains its exponential shape, but the
characteristic time constant $\tau_{obs}$ grows with the afterpulsing probability, as shown in Fig.~\ref{fig:afterpulsing_time_constant}.

The value of  $\tau_{obs}$ is a relevant parameter to describe the performance of a photodetector at the given operating
conditions but it does not, in general, reflect the time constant of the underlying physics process,  $\tau_{aft}$.
The latter can be
determined better by studying the time distribution of the 'first' afterpulse. Such a method would yield a correct result in the
case when the number of afterpulses to a given pulse is one. A Poisson nature of the number of produced afterpulses leads, however, to a systematic
underestimate of
the   $\tau_{aft}$ as shown in Fig.~\ref{fig:first_afterpulse}.

\section{Modeling of  Afterpulses; Multiple Thermal Pulses} \label{model_multi}

'Dark' pulses, i.e. pulses in the absence of the light signal consist of several components. The primary
source of these pulses are the thermal excitations of charge carriers leading to the avalanche process with the $P_{Geiger}$
 probability. The rate of these pulses depends on the temperature and the bias voltage, but the resulting pulses are randomly
distributed in time with the overall rate
\begin{equation} \label{eq:rate}
R(T,V_{ov}) = R_{thermal}(T) \times P_{Geiger}(V_{ov})
\end{equation}
where $R_{thermal}$ is a rate of thermally generated carriers at the temperature T.

 The afterpulses (including the subsequent after-afterpulses) will increase the average rate of the observed pulses by a factor
$1+N_{aft}$ but the time structure of the resulting ensemble of pulses will be depend on the interplay between the afterpulsing
time constant and the typical time distance between the thermal excitations $\Delta t_{thermal}$. The dark pulses will be
distributed randomly in time when  $\Delta t_{thermal} \ll  \tau_{aft}$ or there will be randomly distributed  trains of pulses
(with the train multiplicity and duration dependent on $P_{aft}$)   when  $\Delta t_{thermal} \gg  \tau_{aft}$.

\begin{figure}[ht]
      \begin{center}
      \includegraphics[clip, width=0.5\linewidth]{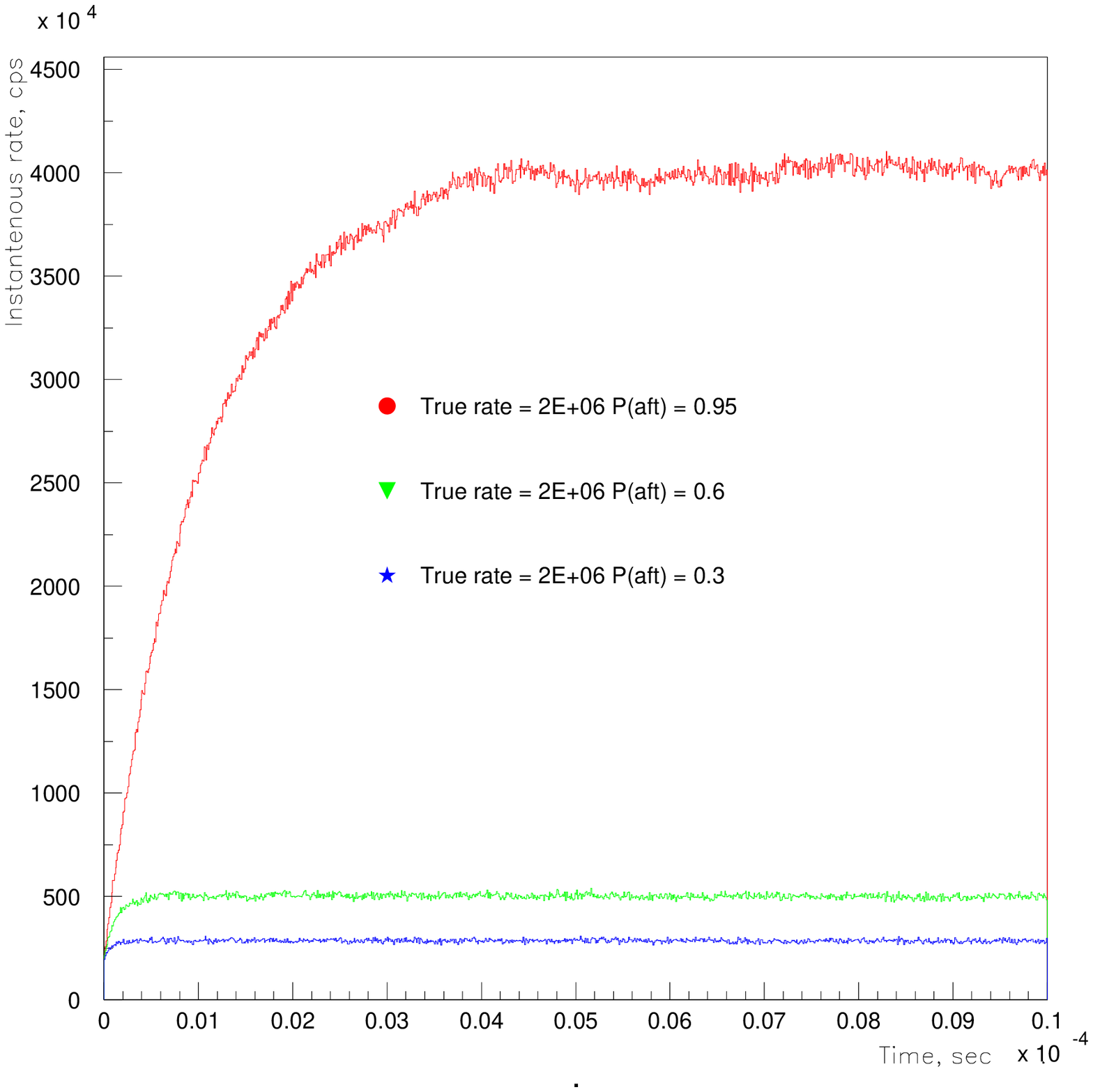}  
      \caption{Instantaneous rate  as function of time after a 'turn-on' for different afterpulsing probabilities. The thermal random rate is $2x10^6 cps$ and  afterpulsing time constant is $50nsec$.}
      \label{fig:instantaneous_rate}
      \end{center}
\end{figure}

Thermally produced pulses have been generated in the interval $0-10\mu sec$  and the resulting afterpulses have been
generated with the average multiplicity given by $P_{aft}$ and time constant  $\tau_{aft}=50nsec$. Fig.
\ref{fig:instantaneous_rate} shows an example of the time distributions of pulses simulated with the thermal random rate of $2MHz$
 and different afterpulsing probabilities. The apparent raise of the instantaneous rate is a reflection of the lack of afterpulses
corresponding to thermal pulses produced at $t<0$. the time interval required to reach the asymptotic level in such a situation
depends on the  interplay of the afterpulsing time constant and the afterpulsing probability.

\section{Thermal Pulses Rate Estimates} \label{mrates}

In the presence of afterpulses the observed 'dark' pulse rate is an overestimate of the rate of random thermal excitations by a
factor $1+N_{aft}$ which can exceed an order of magnitude. A better estimate of the initial thermal rate can be obtained from a
distribution of pulses multiplicity in some time interval $\Delta t$. The probability of observation of zero pulses $P(0)$ in
the given
time interval provides an estimate of the average number of pulses expected in this time interval
\begin{equation} \label{eq:poisson_mean}
<N> = -logP(0)
\end{equation}
If there were no thermal pulses in the specific time interval then there will be no afterpulses either, therefore such method appears
to be systematically superior to the use of the raw observed rate. A possible presence of the afterpulse to the pulses
preceding the time interval in question reduces the probability of observing zero pulses and biases the resulting $<N>$ towards
the higher values, however. To study the resulting bias the random thermal pulses corresponding to the rate of $2x10^6 cps$ were generated in the time interval of $10\mu sec$
with afterpulses corresponding to the time constant $\tau _{aft}$ of $50nsec$.
To avoid an additional bias due to the imperfection of the simulation at the beginning of the
simulated time interval the method was applied using variable gate length, all of the gates starting at $t=7.5 \mu sec$, well
inside the steady state even for large afterpulsing probability, see Fig. \ref{fig:instantaneous_rate}.

\begin{figure}[ht]
\begin{center}
   \begin{minipage}[h]{0.4\linewidth}
      \includegraphics[clip, width=\linewidth]{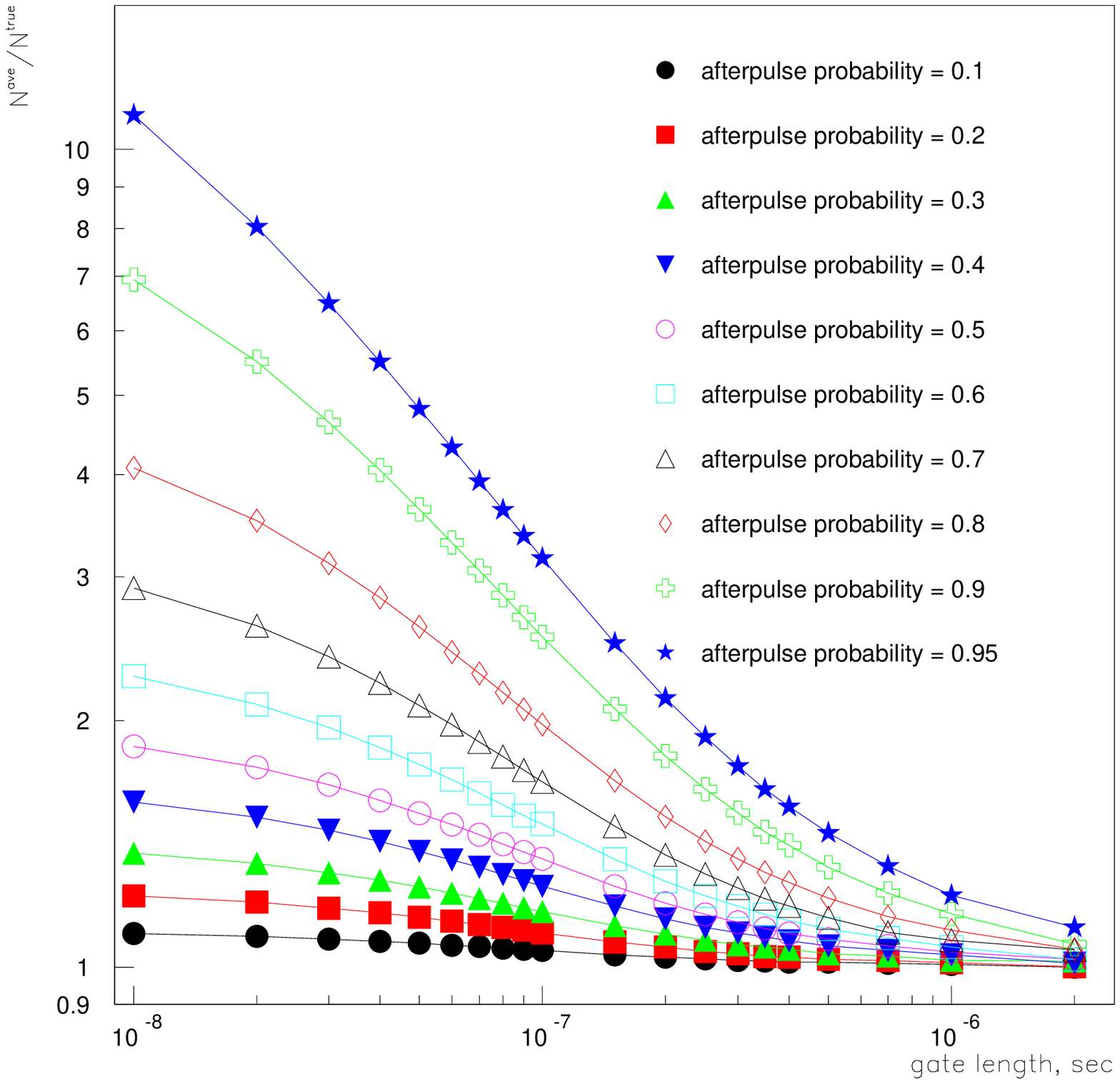}
      \caption{Ratio of the rate of thermal pulses estimated from the $P(0)$ to the true rate of the underlying thermal pulses for different afterpulsing probability $P_{aft}$ as a function of the length of the time interval used.}
      \label{fig:rate_vs_gate}
   \end{minipage}
   \quad
   \begin{minipage}[h]{0.4\linewidth}
      \includegraphics[clip, width=\linewidth]{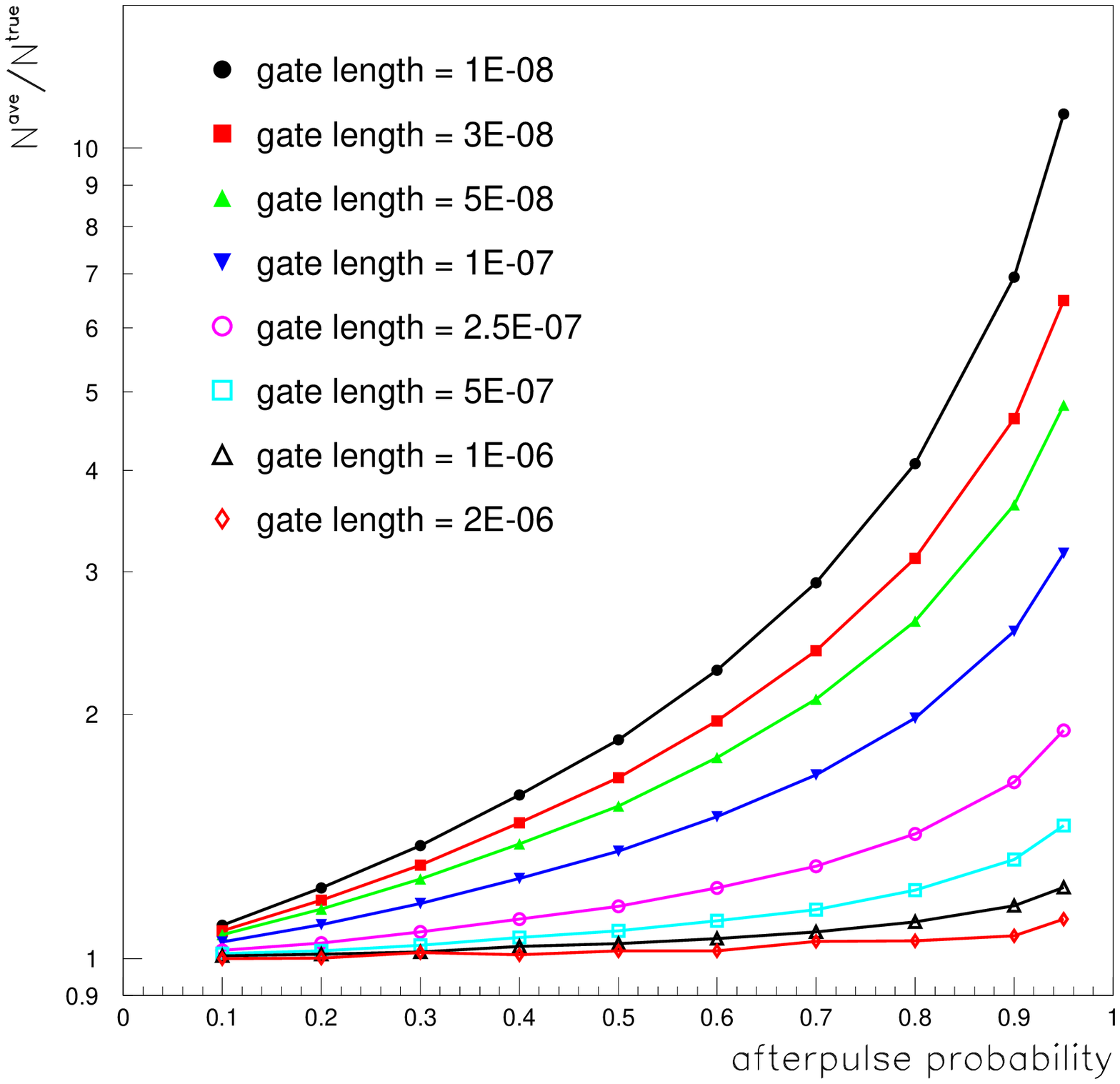}
      \caption{Ratio of the rate of thermal pulses estimated from the $P(0)$ to the true rate of the underlying thermal pulses for different time intervals as a function of  afterpulsing probability.}
         \label{fig:rate_va_paft}
   \end{minipage}
\end{center}
\end{figure}

Figs. \ref{fig:rate_vs_gate} and  \ref{fig:rate_va_paft} illustrate the magnitude of the systematic bias introduced by
afterpulsing in the determination of the thermal pulses rates as a function of the of used gate length and the afterpulsing
probability. For the gate length longer than $~1\mu sec=20\tau_{aft}$ the true thermal rate can be determined with the accuracy
better than
$20\%$ even for very large afterpulsing probability.

\section{Afterpulsing and Absolute Calibration}
\begin{figure}[h]
      \begin{center}
      \includegraphics[clip, width=0.5\linewidth]{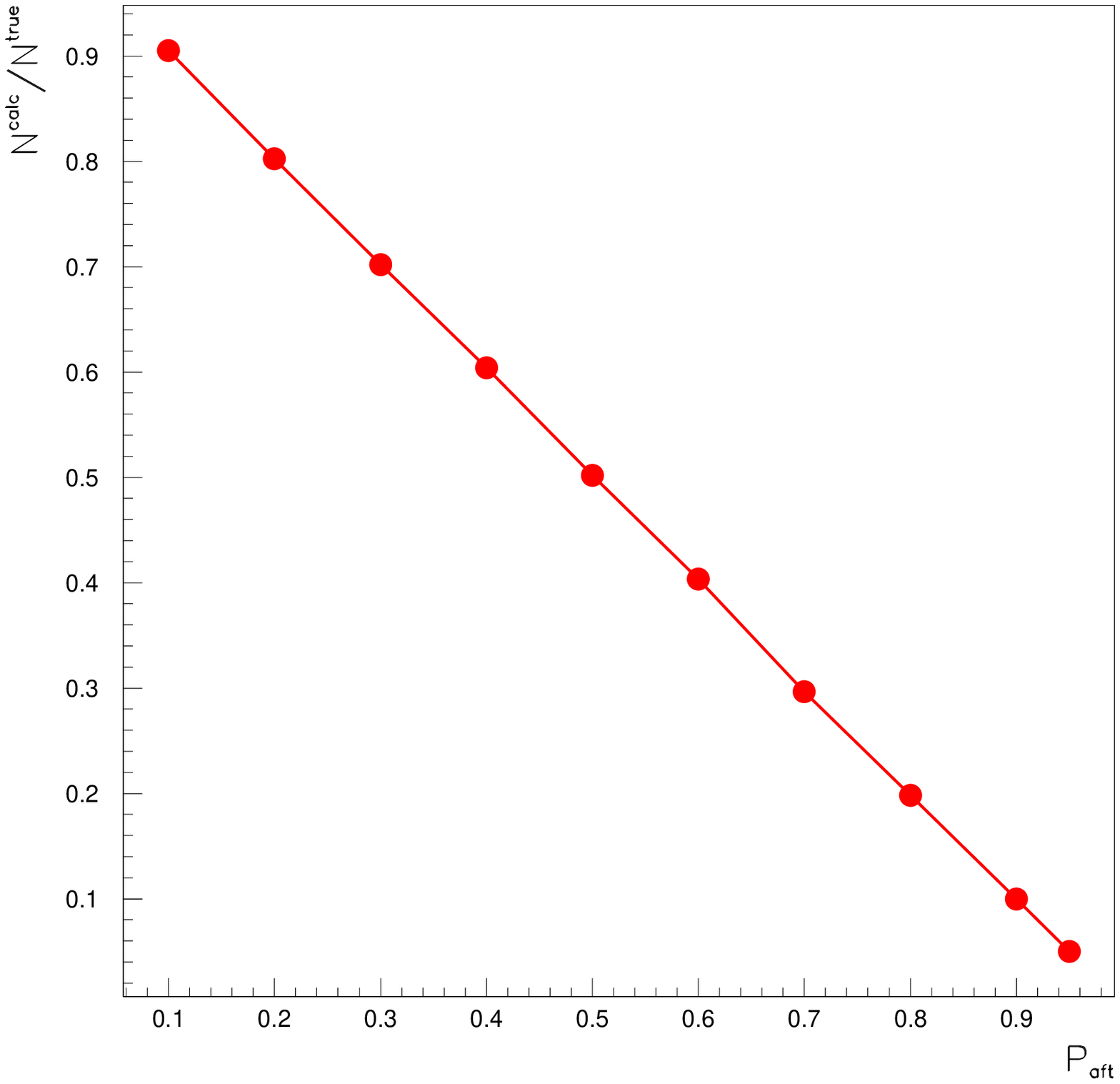}
      \caption{Reduction of size of the photon signal using the Eg.\ref{eq:absolute_cal} with respect to the truth as a function of Afterpulsing probability.}
      \label{fig:n_equiv}
      \end{center}
\end{figure}

Absolute calibration of photodetectors, i.e. conversion of the observed signals into the number of photons impacting the photodetector, is the
crucial step in several classes of experiments. One of they methods successfully used with photomultipliers
involves illumination of the photodetector with a  light signals of constant intensity and determination of the average
number of photons impacting the photodetector from the width of the observed pulse height distribution using the
Eq.~\ref{eq:absolute_cal}
\begin{equation} \label{eq:absolute_cal}
<N> = \frac{Mean^2}{RMS^2}
\end{equation}
where $Mean$ is the mean value of the measured distribution of signals and $RMS$ is the width of this distribution.
This method  relies on the assumption that the variation of the observed signal is dominated by the Poisson fluctuations
of the number of detected photons. It is a very good approximation for the PMT tubes, which have $ENF$ very close to one.
Fluctuations of afterpulsing contribute significantly (Fig.~\ref{fig:ENF}) to the fluctuations of the observed signals thus
reducing the 'determined' number of the incident photons with respect to the true size of the photon signal. This effect can be
parameterized as
\begin{equation} \label{eq:n_equiv}
<N^{det}> = <N^{true}>(1-P_{aft})
\end{equation}
where $<N^{det}>$ is the number of incident photons determined using this method whereas  $<N^{true}>$ denotes the  number of photons
and it is illustrated in Fig. \ref{fig:n_equiv}.

\section{Summary} \label{summary}
Afterpulsing is an intrinsic feature of SiPM-type photodetectors and it does contribute to the various factors relevant to the
performance of these photodetectors. The afterpulsing probability is proportional to the overvoltage, hence many of the observed 
characteristics of the SiPMs depend on the bias voltage. 

A simplified model of afterpulsing gives and explanation of a rapid increase of dark count rate (or a dark current) with the bias 
voltage and, at the same time, it predicts a serious deterioration of the intrinsic resolution of the detectors. The observable
 timing characteristics of SiPMs depend on the bias voltage as longer and longer trains of afterpulses  are generated with the increasing voltage. Various methods of the experimental determination of the true underlying dark count rates and/or absolute
 calibration are affected by the afterpulsing and therefore their outcome  depend on the operating point of the devices. Knowledege of afterpulsing level can be used to derive the corrections to determine the underlying values of these parameters.

 Proper inclusion of the afterpulsing in the analysis of the data is very important, but it
depends on the properties of the photodetector (including the overall rate of afterpulsing and its time constant as well as the
$RC$ recharge time of the cells) as well as on the details of the experimental analysis (gate length). In the simplified model described above the afterpulsing rate is identical with the rate of thermally induced excitations. In the realistic detector the rate
of afterpulses will be reduced by the combine effect of Geiger probability and the recharge of the cell, therefore the relevant  $P_{aft}$ will be in general much smaller then the thermal excitation rate and the effective time constant of afterpulses will be diffrent from the actual lifetime of trapping centers. Nevetrheless it is expected that using the effective $P_{aft}$ and the corresponding time constant should provide a good description of the real devices and
  the results presented here can
be used as a guidance to derive the corrections relevant for specific applications. 
\section{Acknowledgements} \label{acknowledgments}
The author thanks G. Purdue and P. Rubinov for their thoughtful comments and illuminating discussions.
Fermi National Accelerator Laboratory is operated by Fermi Research Alliance, LLC under Contract No. De-AC02-07CH11359 with the United States Department of Energy.





\bibliographystyle{elsarticle-num}



\end{document}